\documentclass[twocolumn,showpacs,prl]{revtex4}

\usepackage{graphicx}%
\usepackage{dcolumn}
\usepackage{amsmath}
\usepackage{latexsym}

\begin {document}

\title
{
Sandpile model on an optimized scale-free network on Euclidean space
}
\author
{
R. Karmakar and S. S. Manna
}
\affiliation
{
Satyendra Nath Bose National Centre for Basic Sciences 
    Block-JD, Sector-III, Salt Lake, Kolkata-700098, India
}

\begin{abstract}
Deterministic sandpile models are studied on a cost optimized 
Barab\'asi-Albert (BA) scale-free network whose nodes are the sites of a square 
lattice. For the optimized BA network, the sandpile model has the same critical
behaviour as the BTW sandpile, whereas for the un-optimized BA network the 
critical behaviour is mean-field like.
\end{abstract}
\maketitle

       Various models of Statistical Mechanics which are usually studied on regular 
   lattices are being studied in recent years on graphs or networks of very complex structures.
   For example, the Ising model has been studied both on the
   Small-World Networks (SWN) \cite {Pekalski,Herrero} and on Scale-Free Networks (SFN) \cite {Aleksiejuk}.
   The phenomenon of Percolation has also been studied on such networks \cite {Newman,Schwartz}.
   In addition disease spreading models like susceptible-infected-susceptible (SIS) \cite {Barthelemy}
   have been studied on networks
   for spreading diseases in the society or spreading of viruses in the Internet.

       Over the last few years it has been observed that the nodal degree 
    distributions of many real-world networks, e.g., World Wide Web \cite {web}
    and the Internet \cite {Faloutsos} are characterized by power law tails:
    $P(k) \sim k^{-\gamma}$ (degree $k$ of a node being the
    number of links attached to it). These networks are called `scale-free networks'
    \cite {barabasi,linked,review,ves} due to the absence of a characteristic value for 
    nodal degrees. Theoretically a number of graphs are generated to model SFNs.
    One of them is by Barab\'asi and Albert (BA) which has the following ingredients, namely:
   (i) A network grows from an initial set of $m_o$ nodes with links connecting all
   $m_o$ nodes. At every time step a new node is introduced
   and is randomly linked to $m (< m_o)$ distinct previous nodes. (ii) Any of these $m$ links
   introduced at time $t$ connects a
   previous node $i$ with an attachment probability $\pi_i(t)$ which is linearly
   proportional to the degree $k_i(t)$ of the $i$-th node at time $t$:
   $\pi_i(t) \propto k_i(t)$.
   For BA model $\gamma=3$ \cite {barabasi}.

       Sandpile models are the prototype models of Self-organized Criticality 
    (SOC) \cite {Bak,Manna,Dhar1,Grass,Man}.
    In these models, long ranged correlations both in space as well as in time
    spontaneously emerge 
    under a self-organizing dynamics, in absence of a fine tuning parameter.
    In its very general form, a sandpile model can be defined on an arbitrary
    connected graph, having a set of vertices connected by another set of edges.
    An integer height variable $h_i$ representing the number of grains in the sand column is associated 
    with every vertex $i$ of the graph. Starting from an arbitrary initial 
    sand height distribution the system is driven by adding unit grains of sand
    at the randomly selected vertices $h_i \to h_i +1$. This sand column is said to be unstable when the 
    height $h_i$ exceeds a pre-assigned threshold value $h_c$. An unstable sand column
    must topple and in a toppling it looses some grains which are distributed among
    the neighboring sites \cite {Bak}. This creates an avalanche of sand column topplings
    and the strength of such activity measures the size of the avalanche.
    There must be some `sinks', i.e., a set of vertices through which grains flow out 
    of the system so that in the steady state the balance of fluxes of
    inflow and outflow currents is maintained.

      In this paper, we have studied the deterministic sandpile model
   on a scale-free network placed on an Euclidean substrate, namely a square lattice.
   The motivation of this study is to acquire support for the validity of our
   recent conjecture \cite {Rumani} that in a sandpile model the precise balance at all lattice sites
   (except on the boundary) between the number
   of outflowing grains $H_i$ which are distributed among the neighbouring sites in a toppling at the site $i$ and the
   number of inflowing grains $H_i'$ received by the site $i$ when its all neighbouring
   sites topple once ensures that the sandpile model behaves like the BTW model \cite {Bak}
   with a multiscaling avalanche size distribution. The absence of the site-to-site balance of
   $H_i =H_i'$ leads to the behaviour of Manna sandpile \cite {Manna}. Below we define and study the 
   sandpile model on the SFN where $H_i$ is equal to the degree $k_i$ of the SFN and therefore
   is an extremely fluctuating quantity. In spite of that the equality $H_i =H_i'$ is maintained
   by construction. We see below that the sandpile model on the optimized SFN indeed behaves
   like the BTW model.

      Recently, BTW sandpile model has been studied on a static model of SFN \cite {Goh}. 
   In contrast to the usual sandpile models there are no specific sinks at fixed positions.
   Instead, during a toppling any grain can evaporate from the system from any
   arbitrary node with a small probability $f$. The distribution of
   avalanche sizes $(s)$ which do not dissipate (i.e., grains do not evaporate in these avalanches) is:
\begin {equation}
{\rm Prob}(s) \sim s^{-\tau}\exp(-s/s_c)
\end {equation}
   where the cut-off of the avalanche size $s_c \sim 1/f$. It is to be noted that the 
   cut-off size does not depend on the network size $N$ but only on the dissipation rate.

       To claim that a dynamical process active in a system is self-organized critical, it is important to 
    ensure that both long ranged spatial and temporal correlations dynamically evolve in this
    system. For the ordinary BTW or Manna sandpiles on systems of spatial extension $L$
    this is verified in the following ways:
    (i) The avalanche size distribution has a power law distribution 
    ${\rm Prob}(s,L) \sim s^{-\tau}$
    for some intermediate range and this range should increase with the system size as the
    cut-off of the avalanche size distribution increases as $s_c(L) \sim L^D$.
    (ii) The average size of the avalanches increases with the system size $L$,
    $\langle s(L) \rangle \sim L^{\nu}$ and $\nu=2$ since the grains while executing a diffusive motion have
    to travel distances of the order of $L$ to go out of the system through sinks situated on the boundary
    Non-zero values of $\nu$ and $D$ indicate
    that system has avalanches of all length scales and the process is indeed critical.

\begin{figure}[top]
\begin{center}
\includegraphics[width=6.5cm]{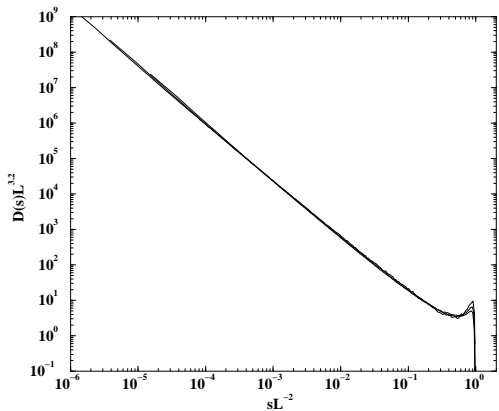}
\end{center}
\caption{
Scaling of the probability distribution of avalanche sizes for the BTW model
on the un-optimized SFN for different
system sizes: $L$ = 256, 512 and 1024. The data collapse gives the
values of the scaling exponents $D = 2$ and $\beta = 3.2$ giving
the avalanche size exponent $\tau_{un-opt} = 1.6$.
}
\end{figure}

       If a sandpile is grown in a closed system (i.e., a system which has no sinks and
    grains do not evaporate from this system) the system eventually reaches a state when an ``infinite avalanche''
    which continues for ever and never stops. Now, if a slow dissipation rate is introduced like every $1/f$
    topplings one grain is dissipated from any arbitrary site of the system, there is no infinite avalanche,
    the system indeed reaches a stationary state, but the avalanche sizes are no more of all length scales.
    This is because the large avalanches loose their strengths by dissipation of grains.
     
       For a network or a graph in general, there is no concept of space, only the connections by links
    between the nodes. One still can define a distance between an arbitrary pair of nodes on a network
    measured by the number of links on the shortest path connecting the two points. The largest of all
    possible shortest paths is called the diameter of the network. A small world network has the
    diameter varying logarithmically with the number of nodes: ${\cal D}(N) \propto \log N$. Since scale-free
    networks are small world networks, it is difficult to observe long ranged spatial correlations
    in sandpile model on SFNs.
   
      Here we study the sandpile model on a SFN constructed on a square lattice
   of size $L \times L$. We first construct a BA SFN of $N=L^2$ nodes. The network starts 
   growing with an initial set of $m_o=(m+1)$ nodes. Each of these nodes is linked to all 
   other $m$ nodes forming a $(m+1)-$clique. After that new nodes are added to the network 
   one by one and each such node is connected to $m$ randomly selected distinct nodes of 
   the already grown network with probability $\pi_i(t)$. This process stops when the 
   network size has grown to $L^2$ nodes. In our calculation we use $m=2$, therefore our 
   network has $L^2$ nodes, $2L^2-3$ links among the nodes and has many loops.
   The nodes of the network are then assigned randomly with uniform probability the sites
   of the square lattice. If two nodes are linked, the corresponding lattice sites are
   connected by straight lines. Thus we place the BA SFN on the square lattice. 
\begin{figure}[top]
\begin{center}
\includegraphics[width=6.5cm]{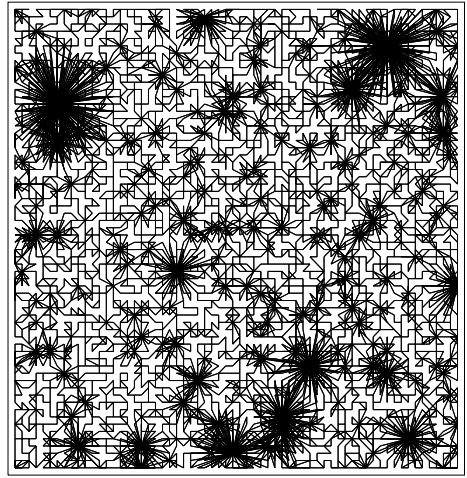}
\end{center}
\caption{
A cost optimized Barab\'asi-Albert scale-free network on a square lattice of
size $L=64$. The SFN of $N=L^2$ nodes is generated by usual BA algorithm whose 
nodes are randomly assigned lattice sites. The cost function ${\cal C}$ 
(the total wiring length) is
then minimized by a large number of trials as described in the text keeping
the nodal degree distribution intact.
Large degree nodes are visible.
}
\end{figure}

      Clearly the degree distribution of such an Euclidean SFN is exactly the same as that
   of the BA SFN. To study the sandpile model we assume that each site (except for sites on the
   boundary) has a site dependent critical height $h^c_i$ of stability which is equal to the degree $k_i$ of 
   the node at that site. Therefore in a toppling, the sand height at site $i$ is reduced to:
   $h_i \to h_i - h^c_i$ and in a deterministic toppling dynamics like BTW model, all the $k_i$ 
   neighbours receive one grain each. The outlet of the
   system is at the boundary. Therefore every boundary site (except the corner sites) has the
   threshold heights $h^c_i = k_i + 1$. This implies that in a toppling at the boundary site
   one grain goes out of the system and never comes back. Similarly at the corner sites
   the threshold heights are $h^c_i = k_i + 2$. Such mechanism of outflow of grains through
   the boundary sites guarantees that the sandpile dynamics on the Euclidean SFN must reach a
   stationary state.
   
      In this Euclidean SFN any site is connected to any other site with equal probability and 
   therefore the average link length $\langle \ell_{ij} \rangle$ is large and of the order of 
   the system size $L$. In a toppling the grains therefore jump large distances on the average. 
   We first study a deterministic sandpile model on such a network. 
   In this sandpile model a grain jumps a distance around $L$ in a toppling.
   Consequently, the spatial extent of all avalanches, small or big, are around $L$.
\begin{figure}[top]
\begin{center}
\includegraphics[width=6.5cm]{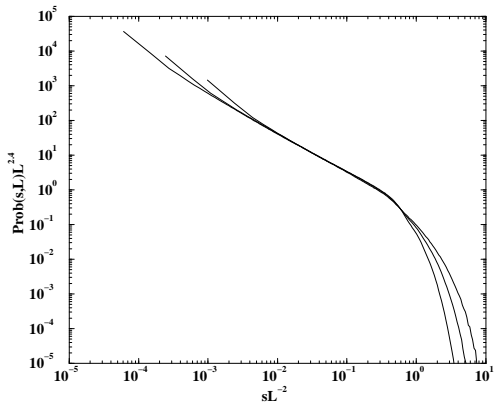}
\end{center}
\caption{
Scaling of data for the probability distribution of avalanche sizes 
for the BTW model on optimized SFN and for different
system sizes: $L$ = 32, 64 and 128. The data collapse gives the
values of the scaling exponents $D = 2$ and $\beta = 2.4$ giving
the avalanche size exponent $\tau_{opt} \approx 1.2$.
}
\end{figure}

       The avalanche size is measured in two ways: (i) the total number of topplings $s$ and the 
   number of distinct sites toppled $a$. A power law distribution of the avalanche sizes with
   a simple finite size scaling implies that the distribution function ${\rm Prob}(s,L)$ obeys
   the following scaling form:
\begin{equation}
{\rm Prob}(s,L) \sim L^{-\beta} f(\frac {s}{L^{D}} ), \quad
\end{equation}
   where the scaling function $f(x) \sim x^{-\tau}$ in the limit of $x \to 0$ giving
   $\tau = \beta/D$, $\tau$ and $D$ are the exponents of the avalanche size distribution.
   One immediate way to check validity of Eqn. (1) is to attempt a data collapse
   by plotting $L^{\beta} {\rm Prob}$ vs. $s/L^D$ with trial values of the exponents.
   It is now well known in the literature that for the Manna stochastic sandpile model
   the distribution obeys FSS with $\tau_{Manna} \approx 1.28$ where as
   for the BTW sandpile the probability distribution ${\rm Prob}(s,L)$ of
   this measure has been found recently to obey a multi-scaling ansatz \cite {Stella1,Stella2}.
   In Fig. 1 we plot the scaling of the avalanche size distribution data for the BTW sandpile
   on the un-optimized BA SFN on the square lattice for $L$ = 256, 512 and 1024. The best collapse
   works for $D=2$ and $\beta=3.2$, giving $\tau_{un-opt}=\beta/D=1.6$. The stochastic Manna sandpile
   is also studied on the un-optimized BA SFN on the square lattice for system sizes upto $L$=1024
   again. We estimated $D=2$ and $\beta=3.0$ giving $\tau_{un_opt}=1.5$. We believe that
   $\tau_{un-opt}$ for both the BTW and Manna sandpiles on the un-optimized BA SFNs are indeed
   mean-field like and both the exponents should be actually 1.5. Similar slight deviation from
   1.5 was also observed in \cite {Goh} for the deterministic case. It is also observed that the
   area $a$ has a similar distribution.

\begin{figure}[top]
\begin{center}
\includegraphics[width=6.5cm]{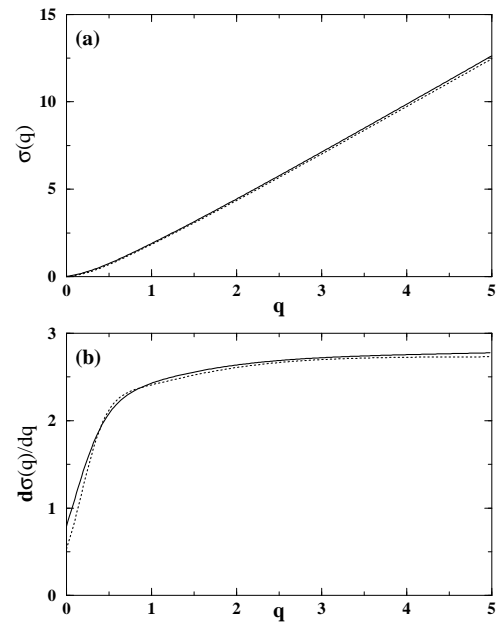}
\end{center}
\caption{
Comparison of the BTW model on ordinary square lattice (solid line) and the BTW model on the
optimized SFN on square lattice (dotted line): (a) the moment exponents $\sigma (q)$ vs. $q$ and
(b) $d\sigma (q)/dq$ vs. $q$.
}
\end{figure}

      Recently a cost optimized SFN on Euclidean space has been constructed \cite {Manna-Alkan}
   where the total sum of the link lengths is optimized keeping the nodal degree distribution 
   exactly same as that of the original SFN. For such a construction one defines a cost function 
   ${\cal C}(N)$ as the total wiring length in terms of the symmetric adjacency matrix ${\bf A}$ 
   of size $N \times N$ (which has elements $a_{ij}=1$ if there is a link between the pair of nodes 
   $i$ and $j$ and 0 otherwise) and the distance $\ell_{ij}$ between nodes $i$ and $j$ as
   ${\cal C}(N) = \Sigma_{i>j} a_{ij}\ell_{ij}$. The optimization process is essentially a rewiring
   process maintaining the nodal degree distribution intact. It starts with a BA SFN constructed
   on a square lattice as mentioned above. A pair of distinct links of the SFN is chosen whose
   nodes are not linked otherwise. One end of each link is then opened and rewired suitably to
   another node of the quartet so that total sum of the rewired length is smaller.
   More precisely, the first node $n_1$ is randomly selected from the set of $N$ nodes and the
   second node $n_2$ is randomly selected from the $k_1$ neighbours of $n_1$. In the same way
   $n_3 (\ne n_1 \ne n_2)$ is selected randomly from $N$ nodes and $n_4 (\ne n_1 \ne n_2)$ is chosen 
   from $k_3$ neighbours of $n_3$. Clearly this move conserves the link numbers as well as degree distribution.
   Rewiring is done following this decision:
   If both $n_1n_3$ and $n_2n_4$ are not linked and also $\ell_{12}+\ell_{34}$ is greater than 
   $\ell_{13}+\ell_{24}$ we link $n_1n_3$ and $n_2n_4$. Another possibility is if
   $n_1n_4$ and $n_2n_3$ are not linked but $\ell_{12}+\ell_{34}$ is greater than $\ell_{14}+\ell_{23}$
   then we link $n_1n_4$ and $n_2n_3$. If both cases are possible we accept one of them with probability 1/2. 
   If only one is
   satisfied we accept that. After rewiring we remove the links $n_1n_2$ and $n_2n_4$.
   If none of the two is satisfied we go for a fresh trial. On repeated trials of these
   moves the cost function gradually decreases. Initially it decreases very fast but eventually
   the success rate becomes very slow. To monitor the optimization process we kept track of the 
   average link length. Our best possible effort yields the average link length 
   $\langle \ell_{ij} \rangle \approx 1.75$ lattice constant. A picture of the optimized
   network is given in Fig. 2. In this best possible optimized
   network the link lenghs ${\ell}$ have an exponential distribution as: ${\cal D}(\ell) \sim \exp(-g\ell)$ with 
   $g \approx 1.16$. Also the diameter of the network
   ${\cal D}(N)$ is measured and is observed to grow as $N^{\mu}$ where $\mu$ is estimated
   to be $0.40 \pm 0.02$. Therefore this network is scale-free but not a small-world network.

The deterministic BTW sandpile model is then studied on such a network.
The avalanche size distribution is calculated for three different system
sizes $L$ = 32, 64 and 128. It was difficult to go beyond this size
because of the large optimization times required. First we tried to make a
scaling plot of the size distribution data. In Fig. 3 we show this plot, 
which shows reasonably well collapse of the data in the intermediate range
of the avalanche sizes. The corresponding $\beta$ and $D$ values fitted are
2.4 and 2 respectively giving a possible value of $\tau_{opt} \approx 1.2$. However
for large avalanche sizes the collapse is much worse and the data for different 
system sizes separate
out. This is a typical behaviour of the BTW like models which show strong
presence of the multiscaling behaviour \cite{Stella1,Stella2}. 

The mutiscaling behaviour is studied in more detail by the evaluation of the
various moments of the avalanche size probability distribution. The $q$-th
moment of the distribution is defined as $\langle s^q \rangle = \int s^q {\rm Prob}(s,L)ds$.
In case the distribution ${\rm Prob}(s,L)$ obeys the finite size scaling
behaviour for the whole range of avalanche sizes, it can be shown that
$\langle s^q \rangle \sim L^{\sigma(q)}$ where
$\sigma(q) = D(q-\tau+1)$ for $q> \tau -1$ and $\sigma(q)=0$ for $0< q <\tau -1$.
The $q$ dependent exponent 
$\sigma(q)$ is determined
from the slope of the plot of $\log \langle s^q (L) \rangle$ with $\log L$, which
in our case are for $L$ = 32, 64 and 128. The interval between successive $q$ values
is 0.02 and moments are calculated at 251 values of $q$ between 0 and 5. In Fig. 4(a)
we show a plot of $\sigma(q)$ vs. $q$ on a linear scale. In Fig. 4(b) the derivative
of $\sigma(q)$ is plotted with $q$. Had the ${\rm Prob}(s,L)$ followed a simple
FSS behaviour the $d\sigma(q)/dq$ in Fig. 4(b) would have saturated for large $q$
values. In stead, the curve gradually increases with $q$, very similar to the
multiscaling behaviour of BTW model. To compare we plot both $\sigma(q)$ and
$d\sigma(q)/dq$ of the ordinary BTW on square lattice studied for same system
sizes with different line styles. We see that in both plots the behaviour is
very similar and the difference between the two curves is very small, within
2-3 $\%$.

      The stochastic Manna sandpile is also studied on the optimized 
   SFNs for small system sizes $L$ 32, 64 and 128. The scaling exponents are estimated as:  $D=2.62$ and $\beta=3.4$ giving
   $\tau_{opt} \approx 1.3$. This value of $\tau$ is compared with the corresponding
   $\tau \approx 1.28$ value of the ordinary Manna sandpile.

      To summarize, we studied the BTW sandpile model on a Barab\'asi-Albert scale-free network
   of $N=L^2$ nodes where the nodes are the sites of a square lattice of size $L$. The SFN
   is then optimized minimizing the total wiring length but keeping the degree distribution
   intact. On such an optimized SFN on the Euclidean space we observe that the sandpile model
   has the same scaling behaviour as the BTW model where as the deterministic sandpile on the
   un-optimized SFN has a mean-field like behaviour.


\end{document}